\documentclass{ws-mpla}
\usepackage[super]{cite}
\usepackage{graphicx}
\begin{document}

\markboth{A. Jakov\'ac, A. Patk\'os}
{Interacting two-particle states in chiral NJL model}

\catchline{}{}{}{}{}

\title{Interacting two-particle states in the symmetric phase of the \\chiral Nambu--Jona-Lasinio model}

\author{\footnotesize A. JAKOV\'AC}

\address{Institute of Physics, E\"otv\"os Lor\'and University, P\'azm\'any P\'eter s\'et\'any 1/A\\
Budapest, H-1117, Hungary \\jakovac@caesar.elte.hu}

\author{A. PATK\'OS}

\address{Institute of Physics, E\"otv\"os Lor\'and University, P\'azm\'any P\'eter s\'et\'any 1/A\\
Budapest, H-1117, Hungary\\
patkos@galaxy.elte.hu
}

\maketitle

\pub{Received (Day Month Year)}{Revised (Day Month Year)}

\begin{abstract}
The renormalisation group flow of the chiral Nambu--Jona-Lasinio (NJL) model with one fermion flavor is mapped out in the symmetric phase with help of the Functional Renormalisation Group (FRG) method using a physically motivated non-local trial effective action. The non-interacting infrared end-point of the flow of the four-fermion couplings is now accompanied by non-zero limiting composite couplings characteristic for interacting two-particle states with finite  energy and physical size. The interaction energy of the constituents is extracted as a function of the physical size of the composite object. The propagation of a two-particle state minimizing the interaction energy has a natural bound state interpretation.    

\end{abstract}

\ccode{PACS Nos.: 11.10Gh,11.10Hi}

\section{Introduction}
Determination of bound state spectra is of central importance in theories of fundamental interactions.  Low energy observables of the physics of strong interactions refer exclusively to bound states, where lattice field theory represents a most successful approach to the first principle determination of the QCD spectra \cite{fodor12,lang16}. Considerable progress has been achieved also with this technique in the non-perturbative investigation of composite Higgs models \cite{lattice-review-of comp-Higgs1,lattice-review-of comp-Higgs2}.  

The Bethe-Salpeter integral equation (BSE) \cite{BSE1,BSE2} is the classic "tool" of quantum field theory in solving the relativistic bound state problem. Its success depends critically on the quality of the kernel-function. Ladder-type resummations of a kernel, involving the exchange of a single force field quantum are the most frequently applied approaches. Finding the appropriate interaction vertex of the constituent field with the force field represents a further challenge. Despite of these limitations, starting from the 1990-ies a rather systematic exploration of the mesonic and barionic bound states has been realized within the BSE-framework combined with Dyson-Schwinger equations for the quark propagators \cite{QCD-BS-article1,QCD-BS-article4, QCD-BS-article2,QCD-BS-article3}. 

At about the same time was also initiated the application of the non-perturbative FRG-approach \cite{wetterich93,morris94} to the determination of bound state excitations of quantum field theories. In Ref. \cite{ellwanger93} the renormalisation group equation (RGE) of the scale-dependent four-point function $\Gamma^{(4)}_k$ has been considered. In the framework of the Wetterich-equation \cite{wetterich93} one writes for its evolution with the scale $k$
\begin{eqnarray}
&
\partial_k&\Gamma^{(4)}_k(p_1,p_2,p_3,p_4)=-\int_q\Gamma^{(4)}_k(p_1,p_2,q,-q-p_1-p_2)\nonumber\\
&&
\times\partial_k\left(G_k((p_1+p_2+q)^2)G_k(q^2)\right)\Gamma^{(4)}_k(-q,q+p_1+p_2,p_3,p_4).
\end{eqnarray}
Here on the right hand side only the partial $k$-derivative of the constituent propagators $G_k(k)$ is taken through the $k$-dependence of its regulator. For the bound state solution of this equation a factorized ansatz has been proposed, whose formal equivalence to the BSE has been demonstrated.

A next step was the proposition for introducing a field $\sigma$ representing the bound state into the effective action by adding a quadratic form to $\Gamma_k$ \cite{ellwanger94}:
\begin{equation}
\Gamma_k[\varphi,\sigma]=\Gamma_k[\varphi]+\frac{1}{2}O^\dagger\star\tilde G\star O-\sigma^\dagger\star O+\frac{1}{2}\sigma^\dagger\star\tilde G^{-1}\star\sigma.
\end{equation}
The field $\sigma$ couples via Yukawa-type convolution (symbolically denoted by $\star$) with the nonlocal field $O(q)$, formed by the constituent fields $\varphi$:
\begin{equation}
O(q)=\int_{p_1}\int_{p_2}g(p_1,p_2)\varphi^\dagger(-p_1)\varphi(p_2)(2\pi)^4\delta(q-p_1-p_2).
\end{equation}
The strategic goal is to find the non-local Yukawa-coupling $g(p_1,p_2)$ and the propagator of the composite field $\tilde G$ from the requirement that the quadratic form  
$\frac{1}{2}O^\dagger\star\tilde G\star O$ maximally cancels the resonant piece developing in $\Gamma_k$ when reaching the scale characteristic for the bound state.

A rather practical "recipe" for the realisation of this strategy has been put forward in 2002 \cite{gies02}. A composite field was introduced into the original action by replacing the pointlike four-fermion interaction by a pointlike Yukawa interaction with the bound state field using a local Hubbard-Stratonovich (HS-) transformation \cite{stratonovich58,hubbard59}. On the example of a gauged chiral NJL model with one fermion species the authors have demonstrated the existence of two (partial) fixed points connected along a one dimensional "renormalized trajectory" following the variation of the effective four-fermion coupling. The neighbourhood of the infrared unstable fixed point represents a strongly coupled interactive theory, which describes via the HS-transformation the dynamics of an elementary scalar field. The other, infrared stable fixed point corresponds to the electrodynamics of massless fermions. It was conjectured that during its evolution towards the infrared the scalar field transmutates into the representative of the positronium-like fermion-antifermion bound state. Since then the extension of any effective theory with fields representing composite (bound) objects with appropriate quantum numbers has become a common usage. The generalisation of the local HS-transformation to larger sets of composite fields has been worked out \cite{floerchinger09,pawlowski07} and important steps were made towards the application of this FRG-technique to $n$-point functions of QCD \cite{pawlowski18}.

The basic steps of the procedure called "dynamic hadronisation" \cite{mitter15,braun16,alkofer18} can be summarized as follows: The composite local field is constructed from the constituent fields with an arbitrary three-variable weight function. Its scale dependence is deduced from the "no-run" requirement enforced on the 4-fermion couplings. Therefore the three-point function has no direct contact to the Bethe-Salpeter wave function of the two-particle composite. The usual initial condition for the RG-equations consists of setting 4-fermion couplings equal to zero at the cut-off. Although the mapping of the 4-fermion theory on the Yukawa-type theory with such a supplementary requirement is in principle exact, but its adequacy is questionable when only a subset of the operators is included into the renormalisation flow.

A construction has been presented recently which ensures the coincidence of the poles in the 2-point functions of the composite fields with the singularities of the 4-point function of the constituent fields by introducing infinite number of auxiliary fields \cite{jakovac18}. This raises
the intriguing question: Can one rely on a unique field representing the complete spectra in a given channel?  What concerns strong interactions, the lowest meson masses might be satisfactorily fitted within a quark-meson effective model with pointlike Yukawa terms \cite{jakovac04,kovacs07,kovacs08}.  Despite of this fact, we cannot know if the s-channel singularity structure of the 4-quark function consistently reproduces them.  The requirement of consistent n-point functions is important also for assessing emergent bound state degrees of freedom in other field theoretical systems.\cite{rose16}

We propose to return to the investigation of non-local extensions of the effective action \cite{ellwanger94}. Parallel progress has been achieved in this direction \cite{dabelow19}, where the fixed point structure of 4-fermion theories with momentum dependent couplings has been explored with functional renormalisation group techniques in 2+1 dimensions. 

The set of couplings whose scale evolution will be traced in the present paper, consists of those which one suspects intuitively to signal bound state formation. In particular, the scale evolution of the 1composite--2fermion vertex function (Yukawa-coupling) will be determined dynamically, not by a constraint equation. Since the resonant state cannot exhaust the spectral function of a certain channel, we trace also the evolution of the pointlike 4-fermion vertices as they fade away in the IR limit. The introduction of auxiliary fields for the spectral eigen-components of the 4-fermion function in the scalar two-fermion sector implements the general strategy of exploring interactive two-particle states developed in our previous paper \cite{jakovac18}.
 
The numerical solution starts in the ultraviolet regime (much above the characteristic momentum scale of the composite state) from the neighbourhood  of the ultraviolet fixed point 
 characterized by a non-vanishhing set of pointlike 4-fermion couplings \cite{aoki97,jaeckel03,braun12}. It will be argued that along some unstable directions the system enters the regime with non-zero Yukawa coupling describing the association of two fermions into a bosonic composite where also a non-zero pointlike 4-point function of the composites exists. 

 One reaches very quickly a scaling regime for the dimensionfull couplings. One finds that the Gaussian fixed point of the 4-fermion NJL theory \cite{braun12,braun17} now attracts in the IR-limit also the Yukawa-coupling and the quartic self-coupling of the composite field. The strength of the  2constituent-1composite 3-point function tends logarithmically to zero as well as the four-boson coupling at an asymptotic rate independent of their starting values.  The truly new feature is that along the trajectories passing in the subspace of the dimensionful composite couplings one observes the existence of a sequence of limiting finite masses and composite object sizes, corresponding to two-particle states of different physical size and well defined interaction energy. These latter features can be tuned smoothly by varying the corresponding initial data of the RG-evolution.

The structure of this paper is the following. In section 2 we extend the effective action of the model with the contributions reflecting the presence of two-particle states composed from two fermions. Via a non-local Hubbard-Stratonovich transformation we arrive at the form of the effective action for which the RG-equations are formulated in section 3. In section 4 the features of the RG flow are analyzed numerically. The results favor the existence of a bound state in the symmetric phase. 

\section{The model and its physically motivated effective action ansatz}

The Fierz complete definining action of the one-flavor chiral NJL-model is the following:
\begin{equation}
\Gamma_{NJL}=\int d^4x\left[i\bar\psi\gamma_m\partial_m\psi+2\lambda_\sigma\left(\bar\psi_L\psi_R\right)\left(\bar\psi_R\psi_L\right)-\frac{1}{2}\lambda_V(\bar\psi\gamma_m\psi)^2\right].
\end{equation}
We shall investigate the emergence of a bound state in the scalar-pseudoscalar channel by explicitly introducing into the scale dependent effective action a four-point term which describes the propagation of a non-local composite two-particle operator  in the $s$-channel. One still keeps non-resonating pointlike four-fermion interactions with strength $\delta\lambda_\sigma, \lambda_V$:
\begin{eqnarray}
&
\Gamma_{NL-NJL}=\int d^4x\left[\bar\psi i\gamma_m\partial_m\psi-\frac{1}{2}\lambda_V(\bar\psi\gamma_m\psi)^2
+2\delta\lambda_\sigma(\bar\psi_R\psi_L)(\bar\psi_L\psi_R)\right]\nonumber\\ 
&
+2\int d^4x\int d^4y\int d^4x_1\int d^4x_2\int d^4y_1\int d^4y_2
~~~~~~~~~~~~~~\nonumber\\
&
\times
\bar\psi_R(x_1)\psi_L(x_2)\Delta_C(x-x_1,x-x_2)G_C(x-y)\Delta_C(y-y_1,y-y_2)\bar\psi_L(y_1)\psi_R(y_2).
\label{NL-NJL-extension}
\end{eqnarray}
The binding function $\Delta_C(x-x_1,x-x_2)$ propagates the $U_A(1)\times U_V(1)$ (globally) invariant fermion and anti-fermion elementary fields from the location $(x_1,x_2)$ to the location $x$ of the composite. The propagation of the composite field is accounted by the propagator $G_C(x-y)$. Both functions have to be determined dynamically.

One could think that keeping the four fermion operators as well as the Yukawa representation of them is double counting, because after the integration of the scalars the four fermion interactions appear anyhow. But, as it was emphasized earlier, we treat the two contributions in a different kinematical regime, and keep the Yukawa representation for the resonant behaviour. In this way the two contributions are clearly distinguishable.

One can easily construct a non-local Hubbard-Stratonovich transform introducing the auxiliary scalar $\Phi_S(x)$ and pseudoscalar $\Phi_5(x)$ fields with the following correspondence to nonlocal composites:
\begin{eqnarray}
&&
\Phi_S(y)\leftrightarrow i\int dx\int dx_1\int dx_2G_C(y-x)\Delta_C(x-x_1,x-x_2)\bar\psi(x_1)\psi(x_2),\nonumber\\
&&
\Phi_5(y)\leftrightarrow \int dx\int dx_1\int dx_2G_C(y-x)\Delta_C(x-x_1,x-x_2)\bar\psi(x_1)\gamma_5\psi(x_2).
\end{eqnarray} 
If one also adds to the effective action a local potential term which is fourth power in the composite fields, then one has
\begin{eqnarray}
&
\Gamma_{NL-NJL}=
\frac{1}{2}\int dx\int dy\left[\Phi_S(x)\Gamma_C^{(2)}(x-y)\Phi_S(y)+\Phi_5(x)\Gamma_C^{(2)}(x-y)\Phi_5(y)\right]\nonumber\\
&
-i\int dx\int dx_1\int dx_2\Delta_C(x-x_1,x-x_2)\left[\Phi_S(x)\bar\psi(x_1)\psi(x_2)-i\Phi_5(x)\bar\psi(x_1)\gamma_5\psi(x_2)\right]\nonumber\\
&
+\int d^4x\left[\bar\psi i\gamma_m\partial_m\psi-\frac{1}{2}\lambda_V(\bar\psi\gamma_m\psi)^2+2\delta\lambda_\sigma(\bar\psi_R\psi_L)(\bar\psi_L\psi)+\frac{\lambda}{24}\left(\Phi_S^2(x)+\Phi_5^2(x)\right)^2\right].
\end{eqnarray}
The two-point function of $\Phi_S$ and $\Phi_5$ is the same inverse function of the chiral propagator:
\begin{equation}
\int dzG_C(x-y)\Gamma_C^{(2)}(y-z)=\delta(x-z).
\end{equation}
In the symmetric phase the scalar and the pseudoscalar spectra are degenerate. We shall use a simplified one-particle form:
\begin{equation}
\Gamma_S^{(2)}(q)=Z_Cq^2+M_C^2= \Gamma_5^{(2)}(q).
\end{equation}
By the above construction this analytic structure necessarily shows up also in the four-fermion function. The parameter $M_C^{-1}$ is a measure of the actual range of the four-fermion interaction. In other words, all fermions should be within this region for the emergence of any composite from this interaction.
  
If the RG-equations display a set of solutions
characterized by a continuous variation of the limiting IR-value of $M_C^2$ then this ansatz with variable initial $M_C^2$ corresponds to  scanning continuously through a
sector of the spectral components in the s-channel of the four-fermion function. A bound state might be signalled by the existence of well distinguishable minimum in the effective squared "mass" as a function of another physical feature to be introduced next. From the above ansatz, however, the two-particle cut is missing, therefore the decay of the possible bound state is not handled by this investigation.

The other $n$-point function which needs explicit modeling is the 3-point function with one bosonic (composite) and two fermion legs. For a first exploratory investigation a Gaussian shape was chosen both in the center-of-mass momentum of the composite and in the relative momentum of the two fermion constituents:
\begin{equation}
\Delta_C(q_1,q_2)=g_Ce^{-\beta(q_1+q_2-Q)^2}e^{-\alpha (q_1-q_2)^2}, 
\label{gaussian-3-point}
\end{equation}
with scale dependent parameters $g_C, \beta,\alpha, Q$. Here $Q$ is the average of the sum of the Euclidean momenta of the two constituents. 

The 3-point function $\Delta(q_1,q_2)$ corresponds to the Bethe-Salpeter wave function \cite{ellwanger93} (see also Ref.\cite{jakovac18}). It is symmetric in its variables since it is constituted by identical fields. The final focus of the present investigation will be on extracting the interaction potential between the constituents, which is dominated by soft momentum fluctuations ($q_1, q_2\approx 0$). Around the origin any positive wave-function is well approximated by Gaussians.

The IR-value of the parameter $\alpha$ can be intepreted as the square of the physical size of the two-particle state, $\alpha(k\approx 0)\sim R_{phys}^2$, since by the uncertainty argument it measures the average of the fluctuation of $(x_1-x_2)^2$ around zero. The same way $\beta$ measures the square of the size of the region where the center of mass coordinate fluctuates relative to the location of the composite. By the above interpretation of $M_C^{-2}$ also both $\alpha$ and $\beta$ should be of that order. Eventually one has to investigate whether in the IR-limit $\alpha(k=0)\equiv\alpha_{phys}$ and $M_C^{-2}(k=0)\equiv M_{phys}^{-2}$ display the expected near equality. We do not attempt to relate rigidly these physical characteristics, just for reducing the dimensionality of the coupling space the widths of the two Gaussians are set equal for any resolution scale: $\beta=\alpha$.

As an alternative line of thinking the starting value of the RG-evolution of $\alpha$ at the cut-off scale, $\alpha(k=\Lambda)$ has another suggestive interpretation: it can be considered a sort of an initial "impact parameter" at which the two fermions enter into interaction with each other. The smaller is its value the more intensive is the interaction. One expects the IR-value of the squared propagator mass $M_{phys}^2$ will depend on the "impact parameter" of the system (e.g. $\alpha(k=\Lambda)$). 

In the next section, the Wetterich-equations will be deduced for the couplings $\delta\lambda_\sigma, \lambda_V, g_C, \alpha, M_C, \lambda$ with a general regulator function. 

\section{The renormalisation group equations}

It is convenient to split the right hand side of the Wetterich equation \cite{wetterich93,morris94} representing the rate of evolution of the effective potential with the scale $k$ into three pieces:
\begin{equation}
\partial_t\Gamma=-\hat\partial_t\textrm{Tr}\log\Gamma^{(2)}_F+\frac{1}{2}\hat\partial_t\textrm{Tr}\log\Gamma_B^{(2)}+\frac{1}{2}\hat\partial_t\textrm{Tr}\log(I-G_B\Gamma_{BF}^{(2)}G_F\Gamma_{FB}^{(2)}).
\label{wetterich-eq}
\end{equation}
where $\Gamma^{(2)}$ refers to the second functional derivative matrix of the effective action, $\Gamma^{(2)}_{B/F}$ represents the purely bosonic/fermionic derivative matrix e.g. $\delta^2\Gamma/\delta\Phi_i\delta\Phi_j$ or $\delta^2\Gamma/\delta\Psi^T\delta\Psi$, while the mixed matrices ($\delta^2\Gamma/\delta\Phi_i\delta\Psi$ or $\delta^2\Gamma/\delta\Psi^T\delta\Phi_i$ are denoted by $\Gamma_{BF}^{(2)}$ or $\Gamma_{FB}^{(2)}$ $(\Psi^T\equiv (\psi^T,\bar\psi)$). The two-point functions $\Gamma^{(2)}_B, \Gamma^{(2)}_F$ and the propagators $G_B,G_F$ contain a regulator restricting their dynamical variation to momenta above the scale $k$. The variable $t=\ln (k/\Lambda)$ relates the actual scale $k$ to the initial (cut-off) scale $\Lambda$. The differentiation $\hat\partial_t$ is applied exclusively to the $k$-dependence of the regulators.

The first term on the right hand side can be rewritten with help of the massless fermion propagator $G_\psi^{(0)}$:
\begin{equation}
-\textrm{Tr}\log\Gamma^{(2)}_F=-\textrm{Tr}\log\Gamma^{(2)}_F(m_\psi=0)-\textrm{Tr}\log(I+G^{(0)}_\psi\Delta\Gamma_{\bar\psi\psi}).
\end{equation}
The quantitity $\Delta\Gamma_{\bar\psi\psi}$ is independent of the fermi-fields, and depends linearly on the composite fields. Therefore its second derivative contributes a fermion-bubble to the RGE of the composite two-point function 
and its fourth derivative to the running of the pointlike quartic composite coupling (the fermion quadrangle).

The purely bosonic contribution on the right hand side of (\ref{wetterich-eq}) has the following explicit form on constant composite field background:
\begin{equation}
\frac{1}{2}\textrm{Tr}\log\Gamma_B^{(2)}=\frac{1}{2}\int_q\log\left[\left(\Gamma_S+\frac{\lambda}{2}\Phi_S^2+\frac{\lambda}{6}\Phi_5^2\right)\left(\Gamma_5+\frac{\lambda}{2}\Phi_5^2+\frac{\lambda}{6}\Phi_S^2\right)-\frac{\lambda^2}{9}\Phi_S^2\right].
\end{equation}
The second derivative with respect to $\Phi_S$ gives a tadpole contribution to the RGE of $M_C^2$. Since the tadpole is momentum independent, it does not contribute to the field renormalisation. Therefore there is no need to consider the pure bosonic term in non-constant background.
The fourth derivative gives the composite one-loop contribution to the running of itself. 

The RG-equation of the composite scalar two-point function is as follows:
\begin{eqnarray}
&
\partial_t\Gamma_S^{(2)}(P)=
\int_q\Bigl[-\Delta_C(q,-q-P)\Delta_C(-P+q,q)\hat\partial_t\textrm{Tr}_DG_\psi^{(0)}(q)G_\psi^{(0)}(q-P)\nonumber\\
&
+\frac{\lambda}{2}\hat\partial_t\left(G_S(q)+\frac{1}{3}G_5(q)\right)\Bigr].
\end{eqnarray}
The RG-equation of $M_C^2$ is arrived by setting $P=0$ on both sides, while $\eta_C=-\partial_t\ln Z_C$ is found by taking first the second derivative $\partial^2\Gamma_S^{(2)}(P)/\partial P_m^2$ and setting $P=0$ in the derivative. 

The RGE of $\lambda$ is the sum of the pure composite and the pure fermion loop:
\begin{equation}
\partial_t\lambda=24\int_q\hat\partial_t\frac{1}{Z^4_\psi(1+r_F(q))^4q^4}\Delta_C^2(q,-q)\Delta_C^2(-q,q)-\frac{3\lambda^2}{2}\int_q\hat\partial_t\left(G_S^2(q)+\frac{1}{9}G_5^2(q)\right).
\end{equation}
 Here the function $r_F(q)$ regularising the fermion-propagator appears explicitly, while the regulator-dependence of the composite propagators remains hidden.

Since one needs the anomalous scaling of the fermion propagator, also the expression of the fermionic self-energy contribution is found from the third term of (\ref{wetterich-eq}), when the logarithm is expanded to linear power:
\begin{equation}
\partial_t\Gamma_\psi^{(2)}(P)=\frac{1}{2}\int_q\Delta_C(P,q-P)\Delta_C(-q+P,-P)\hat\partial_t\left[(G_S(q)+G_5(q))G_\psi^{(0)}(q-P)\right].
\end{equation}

There are two contributions to the composite boson - fermion - fermion three-point function. In the first the fermion legs interact via t-channel exchange of the composite field itself, in the second a fermion loop is generated by taking into account the four-fermion interactions:
\begin{eqnarray}
&&
\partial_t\Gamma^{(3)}_{\bar\psi\psi S}(P_1,P_2,P_3)
=\frac{1}{8}\int_q\Delta_C(P_1, q-P_1)\Delta_C(P_1-q, q+P_2)\Delta_C(-q-P_2,P_2)\nonumber\\
&&~~~~~~~~~~~~~~~~~~~~~~~~~~~\times\hat\partial_t\left[\textrm{tr}_DG_\psi^{(0)}(q-P_1)G_\psi^{(0)}(q+P_2)\left(G_5(q)-G_S(q)\right)\right]\nonumber\\
&&
~~~~+(\delta\lambda_\sigma+\lambda_V)\int_q\Delta_C(-P_3-q,q)\hat\partial_t\left(\textrm{tr}_DG_\psi^{(0)}(q)G_\psi^{(0)}(q+P_3)\right).
\end{eqnarray}
The RG-equation of the strength $g_C$ is determined by setting all external momenta zero. The running of the width $\alpha$ of the composite "wave function" is determined first setting $P_1=P_2=-P_3/2\equiv -P/2$ (which is the most probable configuration within a composite formed from two identical fields), next taking the second derivative with respect to $P$ on both sides and eventually setting $P=0$. The dominant contribution to the interaction energy is expected to emerge from exchanging soft momenta. The choice of the point $P=0$ is therefore optimal.

Finally, we need the RG-equations of the two types of pointlike four-fermion couplings. One has three contributions all coming from the third term on the right hand side of (\ref{wetterich-eq}). The exchange of two composite bosons gives
\begin{equation}
-\frac{1}{16}\hat\partial_t\int_q\frac{\Delta_C^2(q,0)\Delta_C^2(-q,0)}{Z_\psi^2(1+r_F(q))^2q^2}[(G_S^2(q)+G_5^2(q))(\bar\psi\gamma_m\psi)^2+
2G_s(q)G_5(q)(\bar\psi\gamma_m\gamma_5\psi)^2].
\label{two-composite-exchange}
\end{equation}
The starting form of the 4-fermion action is arrived at after exploiting Fierz-identity $(\bar\psi\gamma_m\gamma_5\psi)^2=2((\bar\psi\psi)^2-(\bar\psi\gamma_5\psi)^2)+(\bar\psi\gamma_m\psi)^2$.

The second contribution comes from combining the exchange of one composite boson and the pointlike (non-resonant) four-fermion interaction:
\begin{eqnarray}
&&
\frac{1}{4}\hat\partial_t\int_q\frac{\Delta_C(0,q)\Delta_C(0,-q)}{Z_\psi^2(1+r_F(q))^2q^2}\nonumber\\
&&
~~~~~\times\Bigl[((\bar\psi\psi)^2+(\bar\psi\gamma_5\psi)^2)(G_S(q)-G_5(q))(-2\delta\lambda_\sigma+\lambda_V)\nonumber\\
&&
~~~~~~~~+((\bar\psi\psi)^2-(\bar\psi\gamma_5\psi)^2)(G_S(q)+G_5(q))(-\delta\lambda_\sigma+\lambda_V)\nonumber\\
&&
~~~~~~~~-(\bar\psi\gamma_m\psi)^2(G_S(q)+G_5(q))\left(\delta\lambda_\sigma+\lambda_V\right)\Bigr].
\label{one-composite-one-local}
\end{eqnarray}
It is clear that the explicit symmetry breaking in the boson propagators would induce symmetry breaking in the four-fermion couplings. In the following we shall substitute here and in all RG-equations $G_S(q)=G_5(q)$.

The third contribution comes from the fermion bubble of the nonresonant fermion-fermion scattering which formally coincides with FRG contributions of the original NJL model:
\begin{eqnarray}
&\displaystyle
-\hat\partial_t\int_q\frac{2}{Z_\psi^2(1+r_F(q))^2q^2}\Bigl((\bar\psi\psi)^2-(\bar\psi\gamma_5\psi)^2)(\delta\lambda_\sigma^2+4\delta\lambda_\sigma\lambda_V+3\lambda_V^2)
\nonumber\\
&\displaystyle
~~~~~~~~~~~~~~~~~~~~~~~+\frac{1}{2}(\bar\psi\gamma_m\psi)^2(\lambda_V^2+2\delta\lambda_\sigma\lambda_V+\delta\lambda_\sigma^2)\Bigr).
\end{eqnarray}
This expression leads to the RGE of the NJL theory with Fierz-complete pointlike couplings \cite{aoki97,jaeckel03,braun12}.

The renormalisation group equations are rewritten in terms of dimensionless couplings, convenient for the fixed point search:
\begin{equation}
\mu_C^2=\frac{M_C^2}{Z_Ck^2},~~ \tilde g_{Cr}^2=\frac{g_C^2e^{-2\alpha Q^2}}{Z_\psi^2Z_C},~~ \alpha_r=\alpha k^2,
~~ \lambda_r=\frac{\lambda}{Z_C^2},~~ \delta\lambda_{\sigma r}=\frac{k^2\delta\lambda_\sigma}{Z_\psi^2},~~\lambda_{V r}=\frac{k^2\lambda_V}{Z_\psi^2}.
\label{dimless-couplings}
\end{equation}
Dimensional infrared quantities are reconstructed by the formulae:
\begin{equation}
M_C^2(t)=\mu_C^2(t)e^{2t}e^{-\eta_C(t)}\Lambda^2,\qquad \alpha(t)=\alpha_r(t)e^{-2t}\frac{1}{\Lambda^2},
\end{equation}
where $\Lambda$ is the scale where the solution of the RG-equations is starting. It is worth  to emphasize that all dimensionful quantities are given in terms of appropriate powers of $\Lambda$. Therefore in the numerics the solution of the equations starts at $t=0$ and the term "physical" refers to the $t\rightarrow -\infty$ limit of $M_C^2(t)/\Lambda^2$ or $\alpha(t)\Lambda^2$. 
The explicit equations of the RG-flow of the dimensionless quantities with specific choice of the regulator function \cite{litim01} can be consulted in an expanded version of this Letter in Ref.\cite{version1}.

\section{RG-flow near (partial) fixed points}

The extended set of RG-equations admits the same two fixed points in the subspace of pointlike couplings like the pure 4-fermion model with pointlike couplings. 
The fixed points of the Fierz-complete chiral symmetric NJL-model (with all bosonic couplings set zero) are well-known \cite{aoki97,jaeckel03,braun12,braun17}. The flow-equations for the rescaled dimensionless 4-fermion couplings 
 have three fixed points:
the an absolutely IR-attractive trivial $(\delta\lambda_{\sigma r}^*,\lambda_{Vr}^*)= (0,0)$, a partially UV-repulsive, strongly interacting and unstable $(\delta\lambda_{\sigma r}^*,\lambda_{Vr}^*)= (-64\pi^2,32\pi^2)$ and a partially UV-repulsive, strongly interacting and stable fixed point:
\begin{equation}
(\delta\lambda_{\sigma r}^*,\lambda_{Vr}^*)=(6\pi^2,2\pi^2),
\label{strong-stable-fp}
\end{equation}
In the following subsections we shall investigate the RG flow near these stable partial fixed points along the coupling-axes characterising the composite fields.

\subsection{The strongly coupled IR-unstable fixed point}

In the framework of the extended effective action we investigate the behaviour of the composite bosonic couplings around the fixed point (\ref{strong-stable-fp}). Since the anomalous dimensions $\eta_C, \eta_\psi$ are both proportional to $\tilde g_{Cr}^2$ they can be neglected in the equation of $\tilde g_{Cr}^2$, and the following linearized equation is valid in the immediate neighbourhood of the fermionic fixed point:
\begin{equation}
\partial_t\tilde g_{Cr}^2\approx -K(\delta\lambda_{\sigma r}^*+\lambda_{Vr}^*)\tilde g_{Cr}^2=-4\pi\tilde g_{C r}^2,
\end{equation} 
with the constant $K$ explicitly computable. The  solution $\tilde g_{C r}^2(k/\Lambda)^{4\pi}=1$ 
means that the Yukawa-coupling is IR-relevant around this fixed point. In proportion with $\tilde g^2_{Cr}$ also $\eta_C$ and $\eta_\psi$ starts to increase. The evolution of $\tilde g^2_{Cr}$ changes over to the weak coupling very fast and the value of $\eta_C$ never exceeds the value $\sim .1$, $\eta_\psi$ is still smaller. 

One notes that as a (small) non-zero initial value of the Yukawa coupling starts to grow around this fixed point, it induces an increase of $\mu_C^2$. The linearized part of its equation is written in this regime as
\begin{equation}
\partial_t\mu_C^2=-2\mu_C^2+\frac{1}{4\pi^2}\tilde g_{Cr}^2,
\end{equation}
and one finds for the solution
\begin{equation}
\mu_C^2(t)=\mu_C^2(t=0)\left[1+\frac{\mu_C^{-2}(t=0)}{8\pi^2(1-2\pi)}\right]^{-1}\left[\left(\frac{k}{\Lambda}\right)^{-2}+\frac{\mu_C^{-2}(t=0)}{8\pi^2(1-2\pi)}\left(\frac{k}{\Lambda}\right)^{-4\pi}\right].
\end{equation}
The four-boson coupling stays zero at linear order.Its equation starts quadratically, therefore its variation is slower than powerlike.

The width of the bosonic wave-function follows the leading order equation
\begin{equation}
\partial_t(\tilde g_{C r}\alpha_r)=2\tilde g_{C r}\alpha_r-\frac{\tilde g_{C r}}{2},
\end{equation}
which implies that $g_ {Cr}\alpha_r\sim Bg_{Cr}$, (with some constant $B$) if it starts with zero at $t=0$. After substitution one finds 
One linearizes the above equation around this fixed value and makes use of the scaling behaviour of $\tilde g_{Cr}$ to find for $\delta\alpha_r=\alpha_r-\alpha_r^*$ the scaling behavior $\sim (k/\Lambda)^{2(1-\pi)}$.

One can conclude that the fixed point (\ref{strong-stable-fp}) is fully IR-unstable along the directions  of the composite couplings $\mu_C^2,\tilde g_{Cr}^2,\alpha_r$.

\subsection{Analytic features of the mass-distorted Gaussian partial fixed point}

Assuming first $\eta_C,\eta_\psi\approx 0$, one instantly finds that the irrelevant nature of the dimensionless pointlike 4-fermion couplings $\delta\lambda_{\sigma r}$ and $\lambda_{Vr}$ is unchanged, and similarly $\alpha_r\rightarrow 0$ for $k\rightarrow 0$.
The rescaled $\mu_C^2$ increases as $\sim k^{-2}$ and its physical value $M_C^2$ remains close to its initial value, only slightly modified due to the interactions. The name "mass distorted Gaussian partial fixed point" is used for the infrared limiting theory of massive non-interacting fields.

In the neigbourhood of the non-interacting partial fixed point, when the scale of the system is driven below the mass scales the RG-flow of $\tilde g_{Cr}^2$ is understood with help of the approximate expression of $\eta_C$ ($\eta_\psi$ becomes very quickly negligible and $\alpha_r\approx 0$ is also a very good approximation):
\begin{equation}
\partial_t \tilde g_{Cr}^2\approx\eta_C\tilde g_{Cr}^2,\qquad \eta_C\approx\frac{\tilde g_{Cr}^2}{4\pi^2},
\label{yukawa-anomdim-asymptotics}
\end{equation}
which leads to the following IR-behavior when starting the solution at $k=k_0$:
\begin{equation}
\tilde g_{Cr}^2(k)=\frac{\tilde g_{Cr}^2(k_0)}{1+\frac{\tilde g_{Cr}^2(k_0)}{4\pi^2}\ln\frac{k_0}{k}}.
\label{logarithmic-ir-decrease}
\end{equation}
It is logarithmically approaching zero, independently of the initial coupling. 
In this way, the anomalous fermionic and bosonic dimensions are both negligible in the immediate neighborhood of the fixed point.

The asymptotic flow of $\lambda_r$ is governed by the following approximate equation, after the  composite bubble diagram is stopped to contribute by the inrease of $\mu_C^2$
\begin{equation}
\partial_t\lambda_r\approx \frac{\tilde g_{Cr}^2}{2\pi^2}\lambda_r-\frac{3\tilde g_{Cr}^4}{\pi^2}.
\end{equation}
Substituting the extreme asymptotic from of (\ref{logarithmic-ir-decrease}) one finds that the inhomogenous $\lambda_r^{inhom}(t)\sim |t|^{-1}$ dominates for $k<<k_0$ over the homogenous solution 
$\lambda_r^{hom}(t)\sim |t|^{-2}$.

\subsection{Detailed numerical study of the flow near the mass distorted  Gaussian partial fixed point}

The RG flow was studied around the Gaussian partial fixed point numerically. Our intention was to set the initial values of the couplings at "macroscopic" distance from the non-interacting fixed point, though much closer than to the strongly coupled fixed point. 
The starting value of $\alpha(t=0)\sim R_{phys}^2(t=0)$ controls the initial size of the system. Very large values (e.g. $\alpha_r=100-1000$) correspond essentially to uniform spatial distribution of the constituents. In this limiting case the effect of the quantum fluctuations on $M_C^2$ should be interpreted as the sum of the self-energies of the separate constituents. One expects that the initial contributions (from the far ultraviolet) should follow the same $t$-dependence when $\alpha_r(0)$ is lowered. This argument suggests to select the initial value of $\mu_C^2$ by requiring the same initial slope of the physical mass as $\alpha_r(0)$ is varied: 
\begin{equation}
\frac{dM_C^2}{dt}(\alpha_r(0),\mu_C^{2}(0))=\kappa. 
\label{derivative-fixing}
\end{equation}
This procedure seems to be somewhat arbitrary, since it appears that the value of the constant $\kappa$ is a second dimensionfull independent parameter in addition to $\Lambda$. Its role is to select the UV equivalent physical conditions (i.e. it could be thought as an independent  renormaliztation condition). The question of the actual number of independent parameters will be discussed and clarified at the end of the section. 

For each set we start at $t=0$ and follow the evolution to $t=-10$ (restricted by the numerical stability of the applied Mathematica procedure). This is more than sufficient to recognize the asymptotic tendencies in all couplings for $k\rightarrow 0$.

The features of the  RG-flow for a generic set of the initial parameters can be summarized as follows. 
\begin{itemlist}
\item{First peek in the dimensionless quantities. If we plot the rescaled Yukawa coupling $Z_C\tilde g_{Cr}^2$ after a transient increase a clear constant asymptotics is seen. This is consistent with and indirectly checks numerically the validity of both asymptotic equations in (\ref{yukawa-anomdim-asymptotics}). In case of the scalar self-coupling $\lambda_r$, we observe only very slow decrease, making hard to draw any quantitative conclusion on its compatibility with the above limiting behavior.} 

\item{The anomalous dimension of the composite scalar field converges to zero in quantitative agreement with the expectation based on the estimated asymptotics in (\ref{yukawa-anomdim-asymptotics}). The anomalous dimension of the fermi-field stays very small from the very start of the RG-evolution.}

\item{ The 4-fermion couplings $\delta\lambda_{\sigma r},\lambda_{Vr}$ are, by their physical dimensions, irrelevant in the IR around the Gaussian fixed point. By rescaling with the canonical dimension (e.g. dividing them by $k^2$) we find an almost constant behavior.  We remark that starting with large enough values these couplings would drive the solution of the RG-equations to instability, which indicates in case of the 4-fermion couplings the onset of the broken symmetry regime.\cite{braun12,braun17}}

\item{Other dimensionful quantities are $M_C^2$ and $\alpha$. We plot in Fig.~\ref{MC-alpha-v} the physical value of $M_C^2$ in units of $\Lambda^2$ and the dimensionless combination $\alpha M_C^2$. As it can be seen from the figures, both quantities converge very quickly to their asymptotic nonzero values. Note that typically, relative to the starting value the quantum fluctuations only minimally change the IR-value of $M_C^2/\Lambda^2$ ($0.5\rightarrow 0.494$ in the present case). If one fixes the value of $M_C^2(k=0)$ to some value in $GeV^2$ then the physical size $R^2_{phys}=\alpha(k=0)$ is determined by the asymptotics as a well-defined value in $GeV^{-2}$.} 
\begin{figure}[htbp]
\begin{center}
\includegraphics[width=2in]{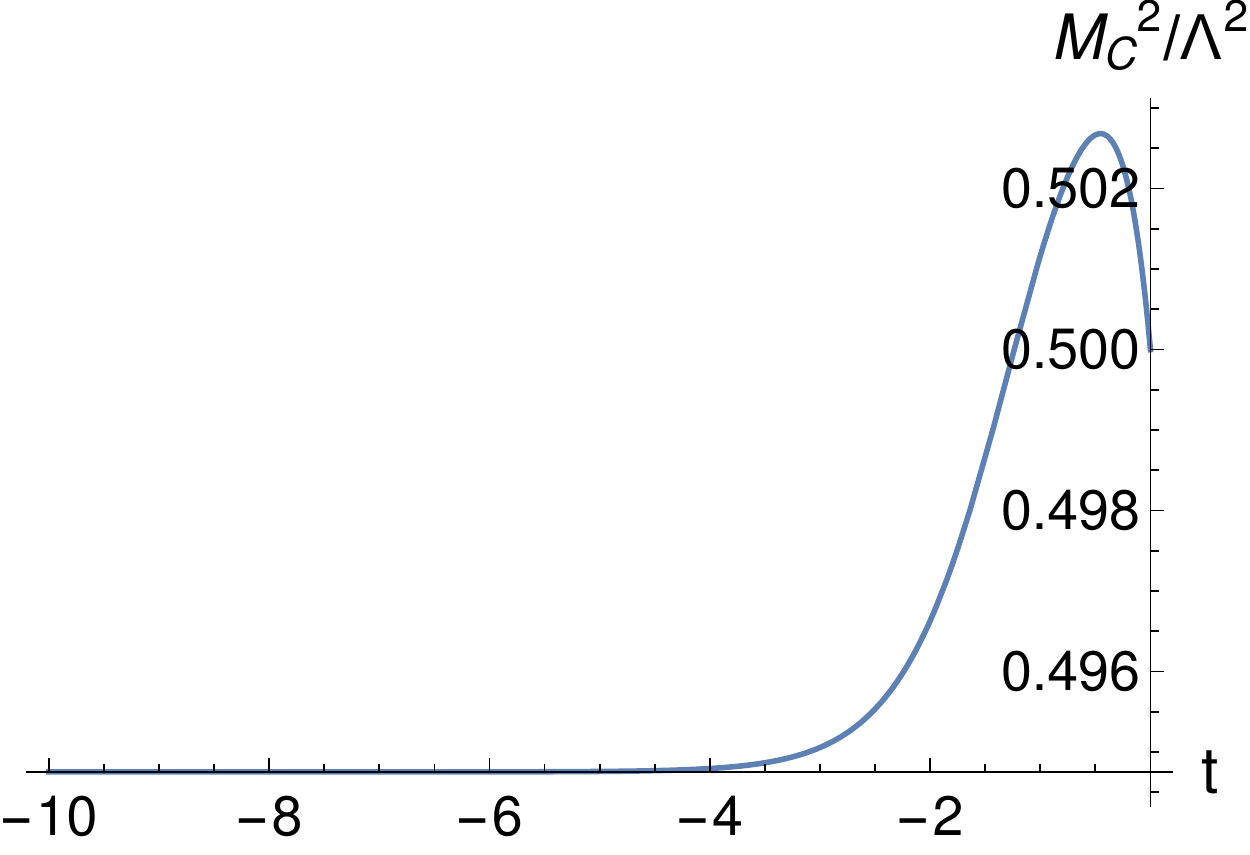}\qquad
\includegraphics[width=2in]{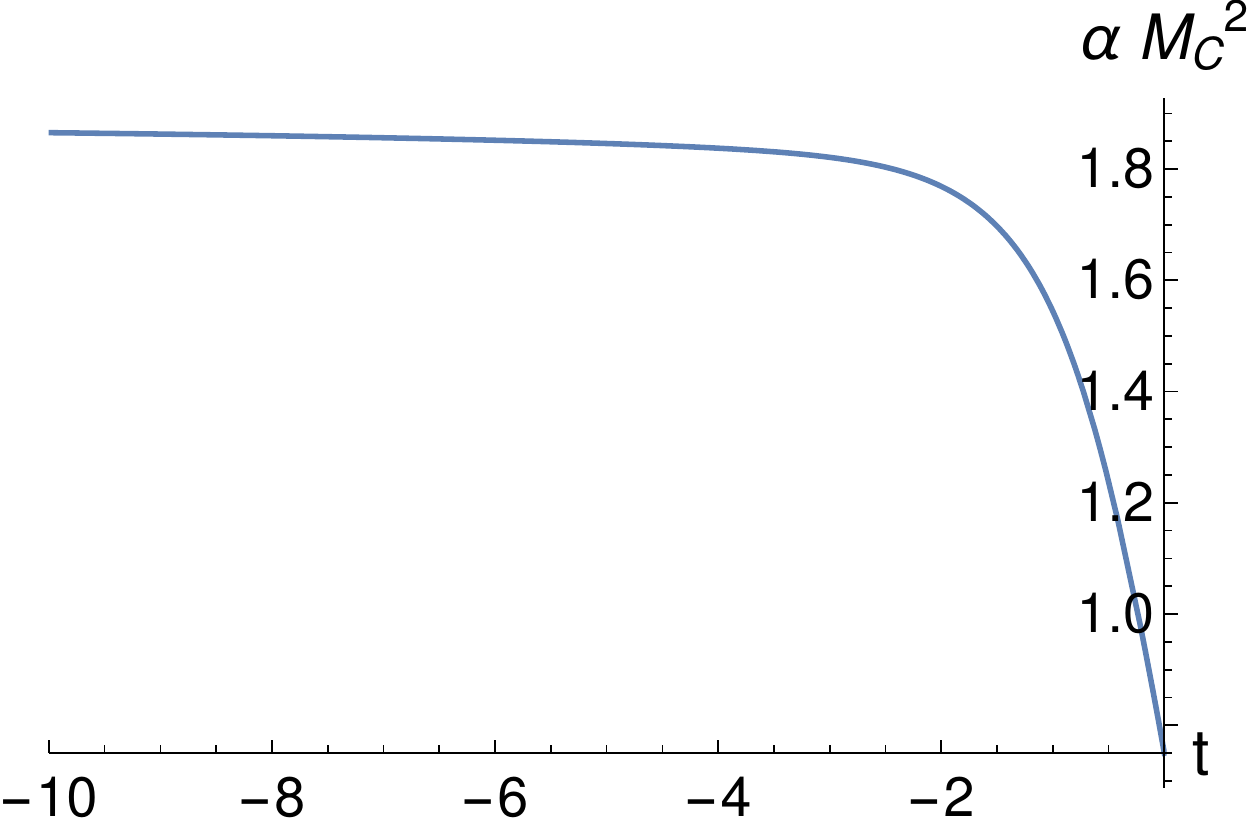}
\end{center}
\caption{RG-variation of the squared boson mass in units of $\Lambda^2$ and of the width of the composite "wave function" in units of $M_C^{-2}$.}
\label{MC-alpha-v}
\end{figure}

\item{Next,  we explore the behavior of the effect of quantum fluctuations on the squared mass parameter by displaying $\delta M_C^2(\alpha(k=\Lambda))\equiv M_C^2(k=0,\alpha(k=\Lambda))-M_C^2(k=\Lambda,\alpha(k=\Lambda))$ in units of $\Lambda^2$ as a function of $t$ for gradually decreasing values of $\alpha(k=\Lambda)$ (accompanied by the choice of $\mu_C^2(0)$ obeying (\ref{derivative-fixing})). In Fig.~\ref{alpha_mphys2_vs_t} one notices the equal slope starting rise (required by Eq.(\ref{derivative-fixing})) is followed by a drop which gets apparently larger as $\alpha_r(k=\Lambda)$ is lowered. The decrease in $\delta M_C^2(\alpha(k=\Lambda)$ relative to the case of the completely unbound particles ($\alpha_r(k=\Lambda)\approx \infty$) lends itself to a physical identification with an attractive interaction energy.  We emphasize that the size of the drop increases very smoothly as $\alpha$ is diminishing.}
\begin{figure}[htbp]
\begin{center}
\includegraphics[width=2in]{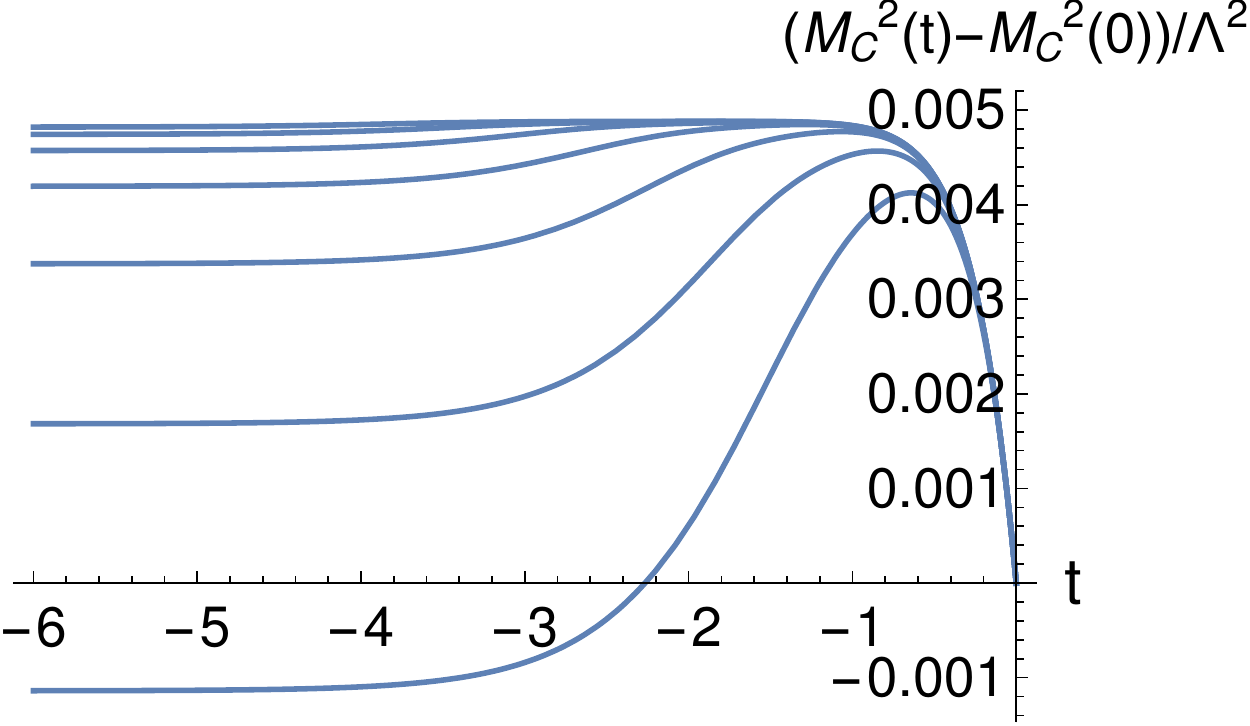}\qquad
\end{center}
\caption{RG-variation of the quantum contribution to the squared composite mass $\delta M_C^2(\alpha(k=\Lambda))=M_C^2(t=-\infty,\alpha(k=\Lambda))-M_C^2(t=0,\alpha(k=\Lambda))$ in units of $\Lambda^2$. From the top to the bottom curves with diminishing $\alpha_r(k=\Lambda)$ are presented.}
\label{alpha_mphys2_vs_t}
\end{figure}

\item{One can display the interaction energies $\Delta M_C^2\equiv\delta M_C^2(\alpha(k=\Lambda))-\delta M_C^2(\alpha(k=\Lambda)=\infty)$ in units of $\Lambda^2$ and the squared physical size $\alpha_{phys}\Lambda^2=\alpha(t=-\infty)\Lambda^2$ as functions of the initial squared physical size $\alpha(k=\Lambda)\Lambda^2$. By the previous figure a sequence of monotonically increasing (less and less negative) interaction energy function is expected with increasing $\alpha(t=0)$. But, in the range $2.5 > \alpha (t=0)\Lambda^2> 1.0$ an opposite tendency takes over (see Fig.~\ref{int-energy-size}), leading to a minimum in the dependence of the interaction energy on the initial size of the two-particle composite. Similar behaviour is observed in $\alpha(t=-\infty)\Lambda^2$ as a function of the initial size.}

\item{The minimum of the final size qualitatively coincides with the maximum of the negative interaction energy. Note when the interaction "energies" are compared among evolutions starting along a different initial slope $\kappa$ the locations of the minima are shifted in a correlated way and also the minimal values change mildly. This observation apparently gives further support to the physical relevance of $\kappa$.}
\begin{figure}[htbp]
\begin{center}
\includegraphics[width=2.4in]{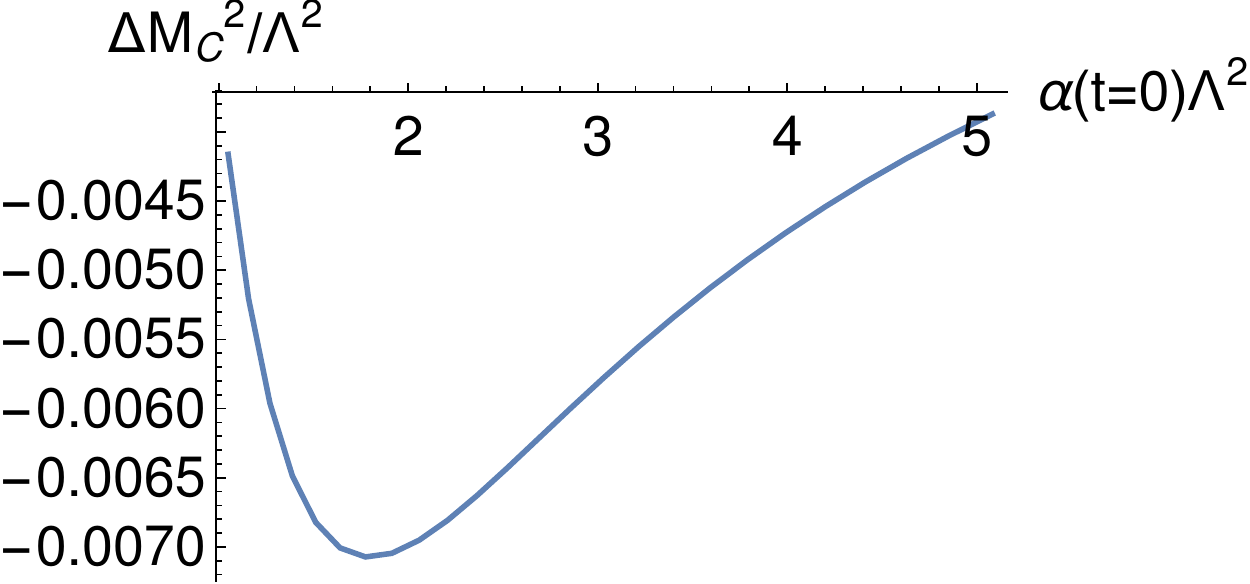}\qquad
\includegraphics[width=2in]{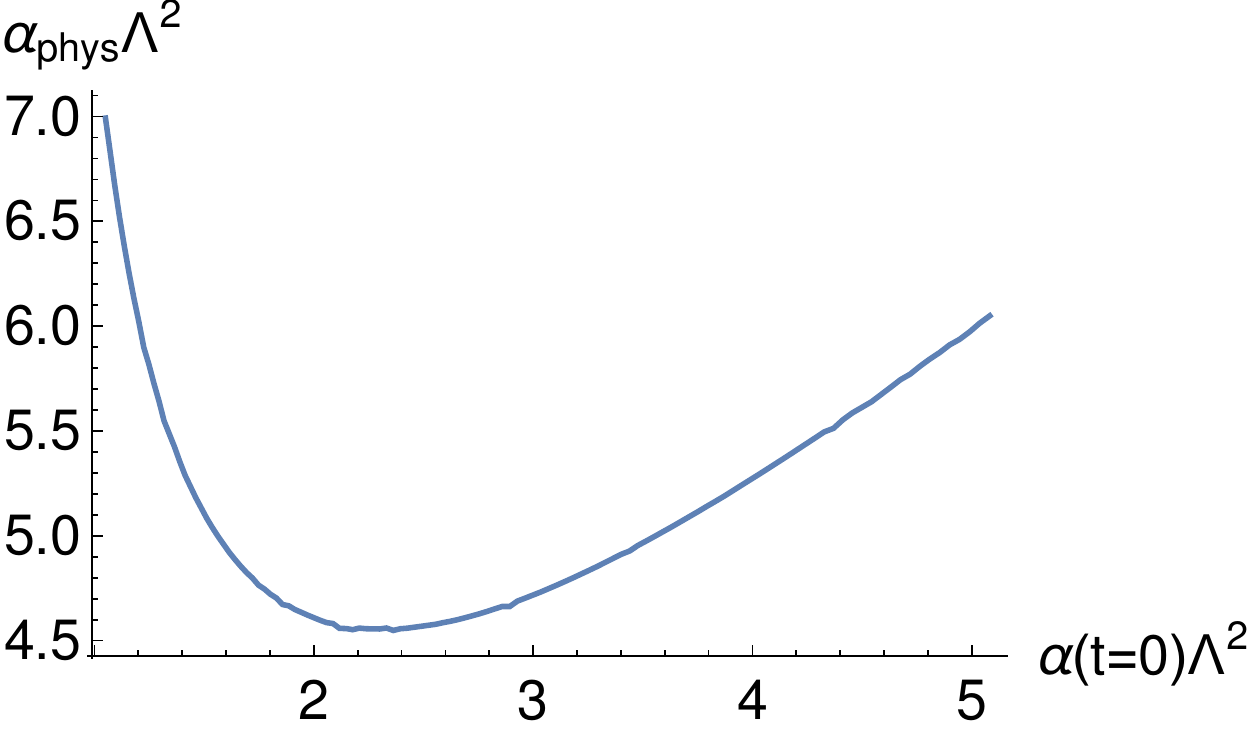}
\end{center}
\caption{Dependence of the interaction energy and of the squared physical size of the composite on $\alpha\Lambda^2=\alpha(k=\Lambda)\Lambda^2$ the cut-off value of the size parameter }
\label{int-energy-size}
\end{figure}
\end{itemlist}

\begin{figure}[htbp]
\begin{center}
\includegraphics[width=2in]{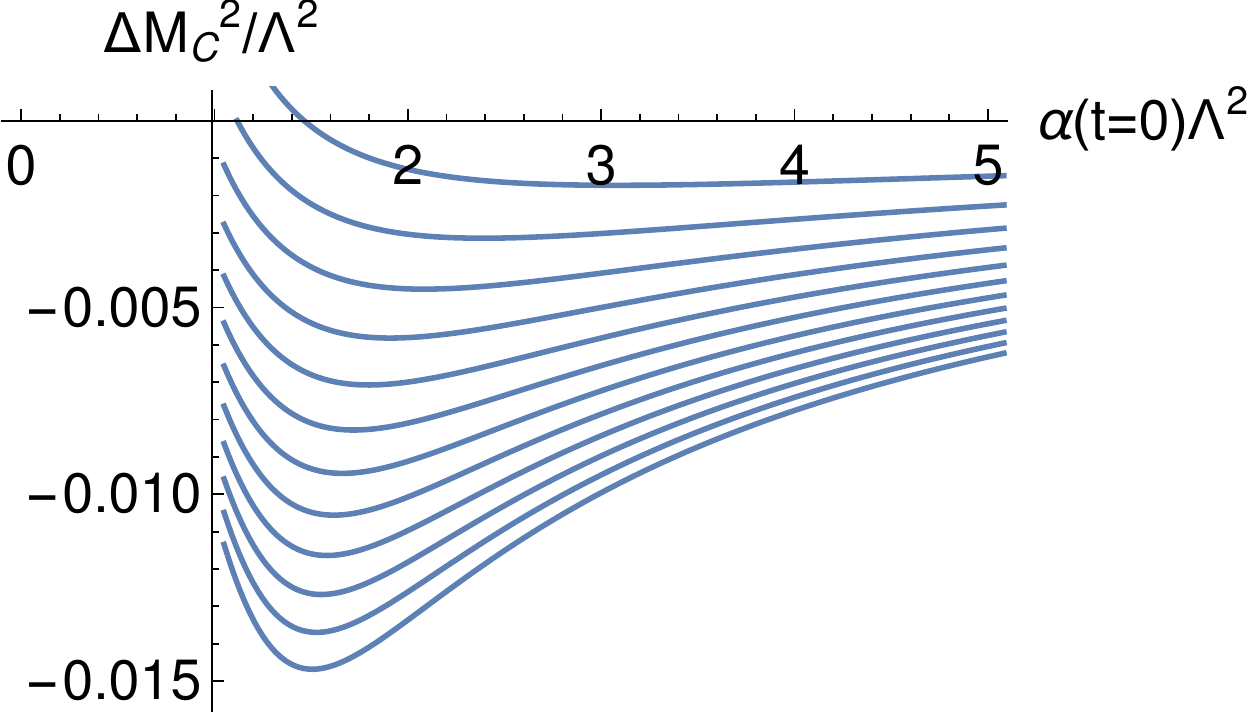}\qquad
\includegraphics[width=2.4in]{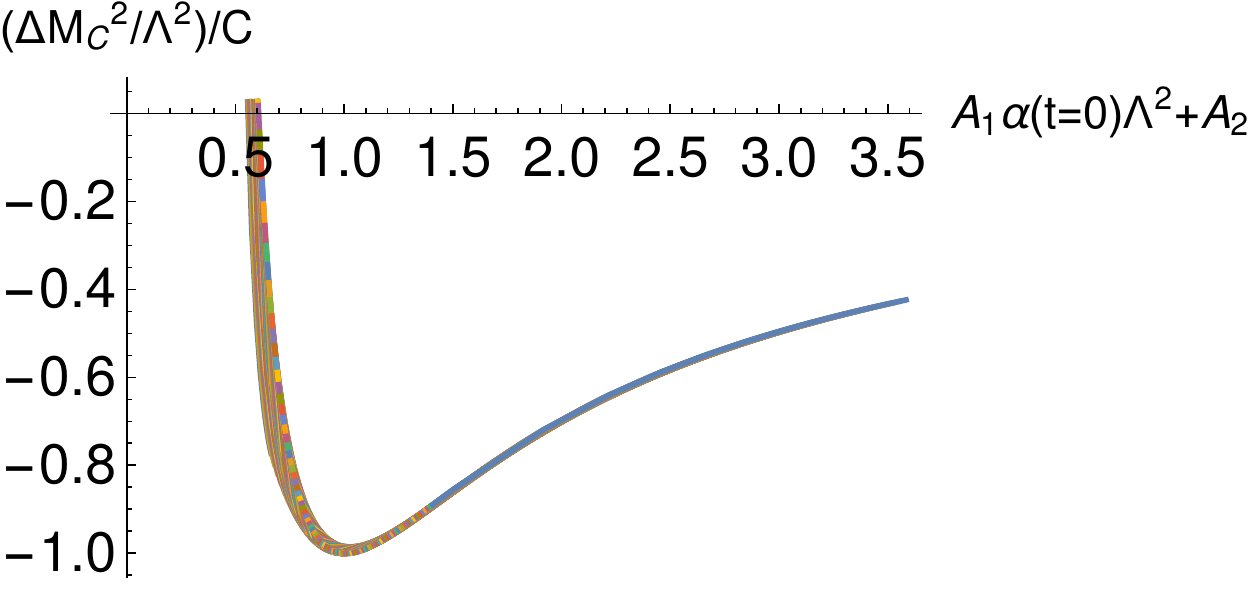}
\end{center}
\vspace*{8pt}
\caption{Left: Dependence of the interaction energy on the input value of the size parameter, for different $dM_C^2/dt$ at $t=0$, Right: the set of potentials scaled together with the mapping $M_C^2/\Lambda^2\rightarrow (M_C^2/\Lambda^2)/C, \alpha(k=\Lambda)\Lambda^2 \rightarrow A_1 \alpha(k=\Lambda)\Lambda^2+A_2$.}
\label{potentials-var-muC2}
\end{figure}

One understands the meaning of the last observation when one searches for a mapping among the "potentials" derived with different starting $dM_C^2/dt$. In the left figure of Fig.\ref{potentials-var-muC2} a set of the potentials is shown for which this derivative has been fixed by choosing at $\alpha_r(t=0)=1000$ the values $\mu_C^2(t=0)=0.08,0.1,0.2,...,1.2$, respectively. Using (\ref{derivative-fixing}) for each of them one finds the corresponding
 $\alpha_r(t=0), \mu_C^2(t=0)$ values and in the same steps as for Fig.\ref{int-energy-size} one constructs the interaction energy vs. initial squared size curve. The smooth deformation due to the change of the initial data allows to map the potentials onto each other (figure on the right). The multiplicative scaling of $\Delta M_C^2/\Lambda^2$ is fixed by the requirement $(\Delta M_{C min}^2/\Lambda^2)/C=-1$. Then the coefficients of the linear mapping of the $\alpha\Lambda^2$-scale $\alpha\Lambda^2\rightarrow A_1(\alpha\Lambda^2)+A_2$ are determined completely by fixing the location of the maximally negative interaction energy to $A_1(\alpha_{min}\Lambda^2)+A_2=1$, which leads with a set of fitted parameters $\{q_{ji}|j=1,2,3;i=1,2\}$ to
\begin{equation}
 A_i=q_{1i}C+q_{2i}+q_{3i}\sqrt{C}.
\end{equation}
In this way the size of the two-particle system with maximally negative interaction energy is determined by the maximal interaction energy (binding energy) itself. The whole curve is now parametrized with help of $|\Delta M_C^2|_{max}/\Lambda^2$. We arrive eventually at the expected conclusion that there is only a single relevant mass-like coupling in this theory.

The most natural emerging interpretation is that the composite configuration which corresponds to the minimum (negative maximum) of the interaction energy is a stable configuration, in other words it is natural to interpret it as a bound state.

\section{Conclusions}

On the basis of the investigations presented in the neighbourhood of the strongly coupled UV and the mass-distorted IR fixed point one might attempt to complement the global RG-flow of the theory restricted to pointlike 4-fermion couplings as displayed in the left figure of Fig.1 in Ref.\cite{braun17}. Our investigation adds several new directions to the $(\delta\lambda_\sigma,\lambda_V)$-plane. Perhaps the most interesting is the flow projected on the 3-dimensional subspace, when $M_C^2/\Lambda^2$ is added. The region studied by us lies below the separatrix in the  $(\delta\lambda_\sigma,\lambda_V)$-plane, where the flow is attracted by the origin. If one starts in this region at $k=\Lambda$ with some initial value of $M_C^2/\Lambda^2$ then the RG-evolution in the 3-dimensional subspace will flow nearly in a plane parallel to the flow in the base-plane, jutst the $t=-\infty$ value of $M_C^2$ will be slightly lowered by the effect of the quantum fluctuations. Similarly would look the projection into the three-dimensional space where the coupling $\alpha\Lambda^2$ is added. On the other hand if the $\tilde g_{Cr}^2,\lambda_r$ couplings are added a universal flow is seen from the strongly coupled to a Gaussian fixed point. It would be very interesting in a next investigation to study the RG-flow of the extended model above the separatrix of the $(\delta\lambda_\sigma,\lambda_V)$-plane towards the broken chiral symmetric phase.

The main new result of the present investigation is to provide a numerical method to extract the interaction energy of the two particles as function of the initial physical size of the composite they constitute. For this we solved the renormalisation group equations derived from an ansatz for the effective quantum action with physically motivated parametrisation. In principle, also in the broken symmetry phase one can extract the interaction potential between the constituents now made massive by the chiral condensate.

It is important to emphasize that by varying the initial width-parameter $\alpha(t=0)$ in the RG-equations we could explore a continuously infinite set of the approximate two-particle eigenmodes of the 4-fermion function of the original model. The resulting squared mass for any single solution in itself was not sufficient to decide whether one deals with a bound-state or only a component of the continuum. Only, by considering the complete set, we were able to single out the state with maximally negative interaction energy, and suggest a bound state interpretation for it. This way of processing is a fully consistent (though approximate) realisation of the strategy which emerged from our general discussion in Ref.\cite{jakovac18}.  

 One might ask to what extent are these results sensitive to the specific choice of the regulator function proposed in Ref.\cite{litim01}. This choice is optimal in the sense that it is the least sensitive to the truncation of the functional form of the effective potential (the gradient expansion)\cite{litim01a,litim11,nandori13}. One expects the same global phase structure and qualitative features of the flow but with quantitative changes due to possible higher order variations in the beta-functions.

Probably the lack of phenomenological motivation explains that we could not find any bound state investigation (for instance Bethe-Salpeter-type calculations) in the symmetric phase to compare with. This gives an extra push for extending the investigations presented in this paper to the broken symmetry phase of the model. Still, one might speculate that at high temperature where the chiral symmetry is necessarily restored bound state formation similar to what was presented here at $T=0$, might take place ensuring the survival of "mesonic" states.

\section*{Acknowledgement}
This research was supported by the Hungarian Research Fund under the contract K104292. The authors are indebted to the Referees for their detailed and valuable reports on the first version of this Letter.

\end{document}